\documentclass[tinyml]{acmart}
\usepackage{float}
\usepackage{adjustbox}
\AtBeginDocument{%
  \providecommand\BibTeX{{%
    \normalfont B\kern-0.5em{\scshape i\kern-0.25em b}\kern-0.8em\TeX}}}

\setcopyright{rightsretained}
\copyrightyear{2022}
\acmYear{2022}

\usepackage{array,multirow,graphicx}
\usepackage{tablefootnote}
\usepackage{multirow, makecell, caption, booktabs, adjustbox, array}

\begin{document}

\title{in-sensor 24 classes HAR under 850 Bytes}

\author{Ahmed.S Benmessaoud}
\email{ahmed.benmessaoud@aol.com}
\affiliation{%
  \institution{Innovation Academy Mila}
  \city{Mila}
  \state{Mila}
  \country{Algeria}
}
\author{Wassim Kezai}
\email{wassim.kezai@oracle.com}
\affiliation{%
  \institution{Oracle Inc}
  \country{Spain}
}
\author{Farida Medjani}
\email{f.medjani@centre-univ-mila.dz}

\affiliation{%
  \institution{Abdelhafid Boussouf University Center}
  \streetaddress{1 Th{\o}rv{\"a}ld Circle}
  \city{Mathematics and their interactions Laboratory}
}

\author{Khalid Bouaita}
\email{khatec@gmail.com}
\affiliation{%
  \institution{Innovation Academy Mila}
  \city{Mila}
  \state{Mila}
  \country{Algeria}
}

\author{Tahar Kezai}
\email{tkezai@gmail.com}
\affiliation{%
  \institution{Innovation Academy Mila}
  \city{Mila}
  \state{Mila}
  \country{Algeria}
}

\renewcommand{\shortauthors}{Trovato and Tobin, et al.}

\begin{abstract}
The year 2023 was a key year for tinyML unleashing a new age of intelligent sensors pushing intelligence from the MCU into the source of the data at the sensor level, enabling them to perform sophisticated algorithms and machine learning models in real-time. This study presents an innovative approach to Human Activity Recognition (HAR) using Intelligent Sensor Processing Units (ISPUs), demonstrating the feasibility of deploying complex machine learning models directly on ultra-constrained sensor hardware. We developed a 24-class HAR model achieving 85\% accuracy while operating within an 850-byte stack memory limit. The model processes accelerometer and gyroscope data in real time, reducing latency, enhancing data privacy, and consuming only 0.5 mA of power. To address memory constraints, we employed incremental class injection and feature optimization techniques, enabling scalability without compromising performance. This work underscores the transformative potential of on-sensor processing for applications in healthcare, predictive maintenance, and smart environments, while introducing a publicly available, diverse HAR dataset for further research. Future efforts will explore advanced compression techniques and broader IoT integration to push the boundaries of TinyML on constrained devices.

\end{abstract}

\keywords{Intelligent Sensors, Time series, TinyML, ISPU, HAR}

\maketitle

\section{Introduction}

As the Internet of Things (IoT) continues to expand, the number of edge devices connected to networks is growing exponentially \cite{1}. These devices generate vast amounts of data that traditionally have been processed in centralized cloud servers. However, cloud-based processing faces significant challenges, including high economic costs, increased latency due to data transmission, and substantial energy consumption \cite{1}. Additionally, the emergence of generative artificial intelligence (AI), which requires substantial computational resources, exacerbates these issues. To mitigate these challenges, there is a shift toward processing data directly at the edge of the network, known as edge computing, which offers advantages such as improved real-time capabilities, enhanced data privacy, and reduced reliance on cloud infrastructures \cite{3}.

Recent advancements in semiconductor manufacturing have led to the development of Intelligent Sensor Processing Units (ISPUs), which integrate small digital signal processing units directly into sensor hardware \cite{8}. These hardware advancements have given birth to a new niche within tiny machine learning (tinyML), enabling on-sensor processing of data and eliminating the need to transmit raw data to microcontrollers (MCUs) or cloud servers. This approach minimizes power consumption, reduces latency, and enhances system efficiency \cite{danilo}.

\begin{figure}[H]
    \centering
    \includegraphics[width=1\linewidth]{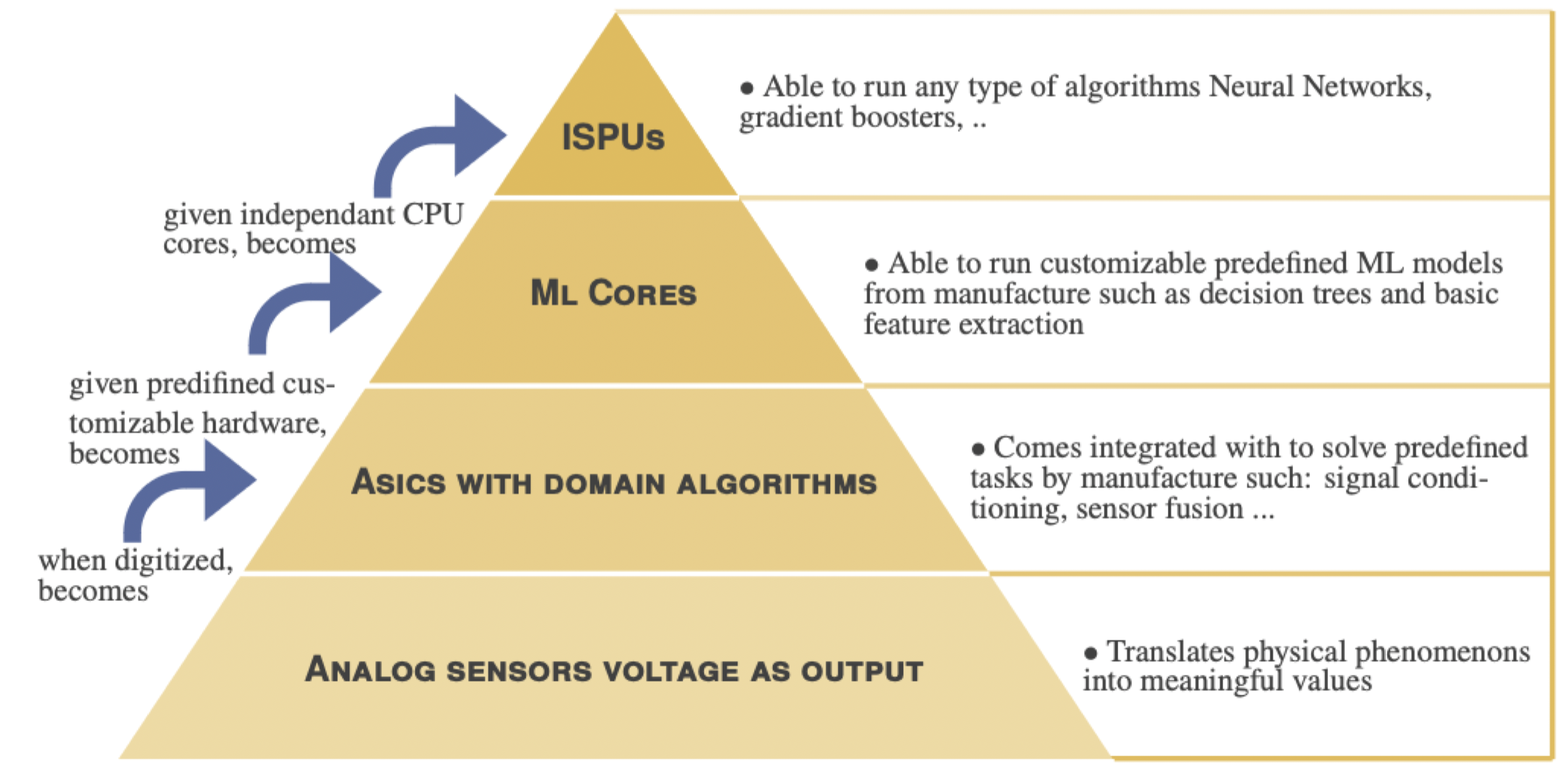}
    \caption{Hierarchical Representation of Evolution of Sensors from Analog to ISPUs.}
    \label{fig:enter-label}
\end{figure}

Deploying machine learning (ML) models directly on ISPUs presents significant challenges due to their extremely limited computational resources, often constrained to memory sizes under 8KB \cite{8}. Traditional neural networks typically require much more memory and computational power, making it difficult to implement them on such ultra-constrained devices. Prior work has focused on model compression and quantization techniques to reduce the size of ML models for deployment on microcontrollers. However, these solutions often still exceed the resource limitations of ISPUs or lead to compromised model performance \cite{in-sensor, 12}.

In this study, we focus on developing a Human Activity Recognition (HAR) model that runs directly on the ISPU core. HAR involves classifying human activities like running, washing face, or using tools based on accelerometer and gyroscope data. It has widespread applications in smart homes, healthcare monitoring, elderly care, bearing Fault detections, and predictive maintenance\cite{hh,pdm, yahia, 2p1, 2p7}. Implementing HAR models directly on sensors enhances real-time processing capabilities, improves data privacy by keeping sensitive information on the device, and reduces dependency on cloud infrastructures \cite{2p1, 2p}. Given the relevance of HAR in IoT and the stringent resource constraints of ISPUs, HAR serves as an ideal application to demonstrate the feasibility of deploying complex ML models on ultra-constrained hardware. We address the challenge of implementing neural networks for HAR on ultra-constrained ISPUs with memory under 8KB. We have developed a novel 24-class HAR model that operates efficiently within these stringent resource constraints. Our model achieves an accuracy of 85\%. Our method utilizes specialized dataset preprocessing and a training pipeline with incremental class injection. This ensures maximum model accuracy and accommodates a larger number of classes, thereby extending the capabilities of the ISPU.\\ Our key contributions are summarized as follows:
\begin{itemize}
    \item Implementing a deep learning model for Human Activity Recognition (HAR) on an Intelligent sensor of  with less than 850 bytes of stack memory usage.
    \item We designed a data preprocessing pipeline for the HAR model. Our training method accurately classifies 24 gestures while keeping memory footprint below 2KB, making it feasible for deployment on limited-resource devices.
    \item Demonstrating the practical implications of on-sensor processing, improving real-time capabilities, enhancing data privacy, reducing power consumption, and scalability.
    \item We release a HAR dataset with 24 distinct gestures, recorded for 12.5 hours. Each gesture is captured by at least 2-5 individuals, promoting diversity and reliability. This resource aids in training and evaluating HAR models, advancing the field.
\end{itemize}

\section{Related Work}

Since the introduction of Intelligent Sensor Processing Units (ISPUs), several studies have demonstrated their potential in ultra-constrained edge computing applications. Mikhaylov et al \cite{r1}, explored the use of the ISM330IS IMU's ISPU for Structural Health Monitoring (SHM) in wind turbine blades, implementing an event-based tinyML model for fault detection. This approach reduced power consumption by up to 75\% and achieved near-zero latency in fault alerting, showcasing the efficiency gains of on-sensor processing.

Similarly, work by Chowdhary et al \cite{in-sensor}, leveraged ISPUs for personalized on-device learning and inference in human activity recognition. Their framework achieved 96.7\% accuracy in activity classification with significant memory optimizations, enabling on-sensor learning under 8 KB of memory. These works highlight the versatility and impact of ISPUs in advancing real-time, energy-efficient edge computing solutions across diverse applications.

\section{ISPU}

\begin{figure}[H]
    \centering
    \includegraphics[width=1\linewidth, trim=0 0 0 1cm, clip]{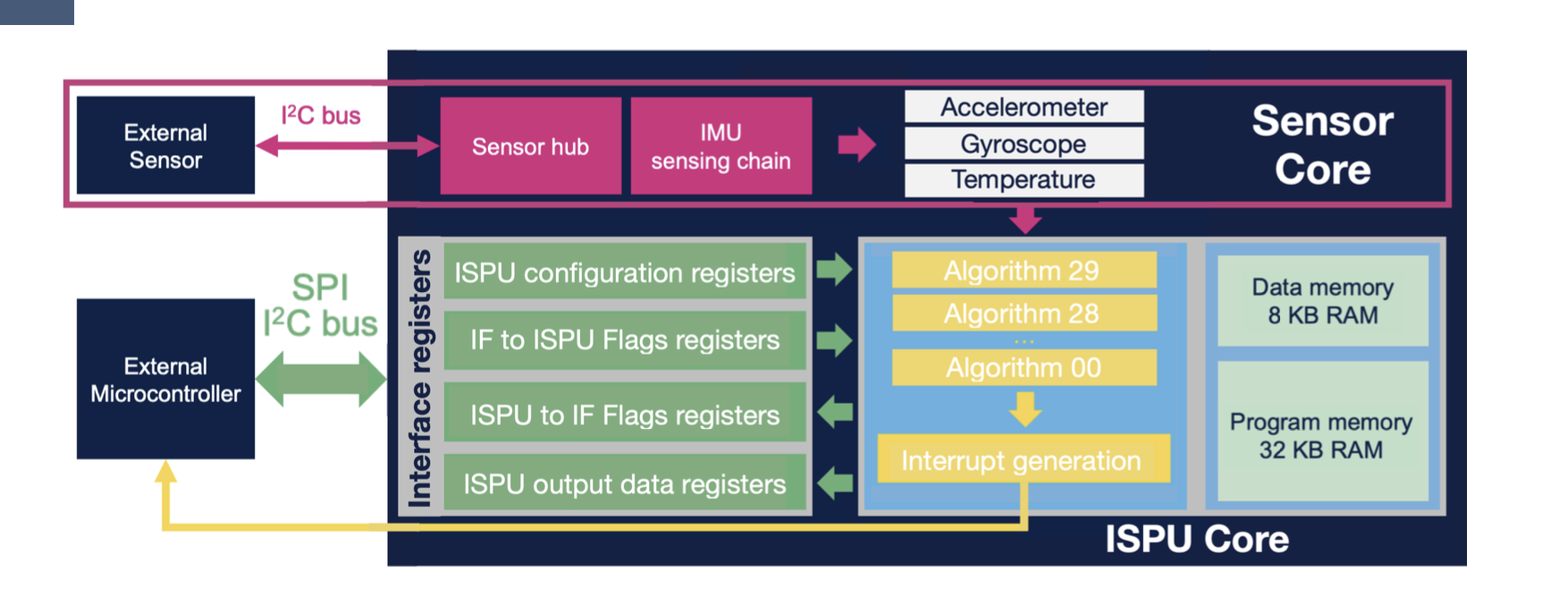}
    \caption{ISPU Chip Architecture}
    \label{fig:ispu}
\end{figure}

The Intelligent Sensor Processing Unit (ISPU) \cite{ism330} is a 32-bit RISC processor designed for ultra-low-power applications, integrated into the LSM6DSO16IS and ISM330IS six-degree-of-freedom (6DoF) sensor ASICs \cite{x6}. Its architecture offers key features that improve performance and efficiency in sensor-based tasks, especially in environments where energy use is a concern. The ISPU consumes as little as 0.46mA, making it ideal for low-power applications. Fig:\ref{fig:ispu} 1 shows the hardware architecture of the ISPU.

Operating at CPU frequencies of 5 MHz or 10 MHz, the ISPU balances processing power with minimal energy consumption. The core is an 8-kilogate Harvard RISC STRED processor that includes a floating point unit, a binary convolutional accelerator, four pipeline stages, and two branch shadows, achieving a CoreMark® score of 70 µW/MHz packaged in LGA-14 \cite{ispu}. 
The ISPU features 8kB of data memory and 32kB of volatile program memory. The ISPU consumes between 0.6mA and 1.2mA at 1.8V and 5–10MHz during normal operation. Compared to a Cortex-M0 core—which consumes 1.6–2.6mA at 1.8V and 4–8MHz—the ISPU uses significantly less power, up to five times less for sensor fusion tasks \cite{ispu}. 
The ISPU is fully programmable in the C language, offering flexibility and ease for developing various machine learning models and processing algorithms. Programs are written in standard C code and compiled using the ISPU toolchain \cite{ispu_toolchain} or X-Cube-ISPU. NanoEdge AI Studio helps developers port machine learning models from popular frameworks like Tensorflow \cite{69}, and XGBoost \cite{xgb} to the ISPU, NeutonAi \cite{neuton} platform allows for the deployment of automl models optimized for STRED architecture.

Each processing cycle involves the ISPU waking up when new data arrives, running the onboard program within the Output Data Rate (ODR) or Interrupt Request (IQR), and then returning to sleep. Using interrupt requests and an interrupt vector table, the ISPU can run up to 30 separate user defined algorithms in which we can implement tinyML models inside them. The ISPU does not have any ROM memory making it requiring another microcontroller to upload the main program to it every time it restarts vis I2C/SPI, this makes it suitable for Master-Slave architecture to manage and distribute processing tasks efficiently between the main unit and subordinate units.

As a matter of fact, deploying neural networks on sensor devices has become more accessible due to simplified deployment. This process involves mainly three steps: processing sensor input, loading pre-trained model weights, and implementing a custom feed-forward pass using C. This ease of deployment enhances the feasibility of embedding sophisticated machine learning capabilities directly onto sensor platforms, advancing TinyML.

\section{Methods}
In this section we will discuss the data preprocessing pipeline steps in sequential order as in Fig \ref{fig:pipeline}:

\begin{figure}[H]
    \centering
        \includegraphics[width=0.8\linewidth]{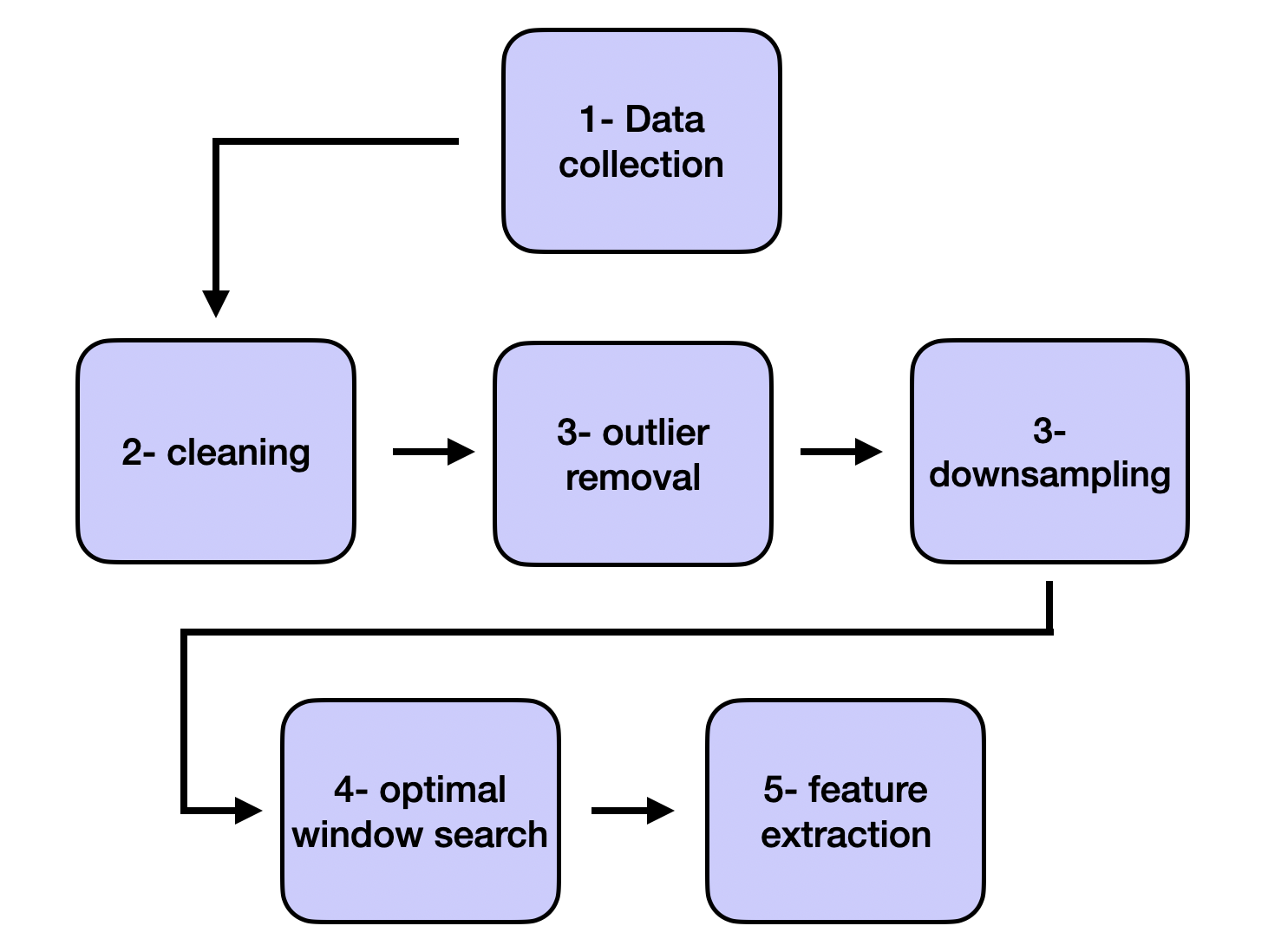}
    \caption{Data Preprocessing Pipeline}
    \label{fig:pipeline}
\end{figure}

\subsection{Data Collection Details}
The data collection process for this study encompassed 31 distinct classes, categorized into groups such as Cleaning, Writing, Sports, Workshop Activities, Kitchen, Unclassified, Miscellaneous, and Other activities. However, as will be explained in Section \ref{inject}, only 24 of these classes will be utilized.

The data collection process for this study encompassed 24 distinct classes, systematically categorized into groups such as Writing, Cleaning, Sports, Workshop, Kitchen, Other, and Unclassified activities. The dataset comprises over 750 minutes of accelerometer and gyroscope recordings, ensuring a comprehensive representation of each class. Data acquisition involved 2 to 5 participants per class, with certain exceptions where the number of participants varied to accommodate specific experimental requirements. To guarantee the precision and reliability of the sensor data: the accelerometer was configured to an 8g range, the gyroscope was calibrated to 2000 degrees per second (dps), and the sampling frequency (ODR) was consistently maintained at 104 Hz.

\subsection{Data Cleansing}

To enhance data quality and maintain the consistency of periodic movements, we employ the following process:
\begin{itemize}
    \item Removed outlier peaks and valleys for accurate feature extraction.
    \item Trimmed noise at the beginning and end of recordings.
\end{itemize}
This process:
These steps were essential for maintaining the overall consistency and reliability of the periodicity of the signals of the dataset, thereby providing a robust foundation for subsequent analysis and the development of effective machine learning models.

\begin{figure}[H]
    \centering
    \includegraphics[width=0.8\linewidth]{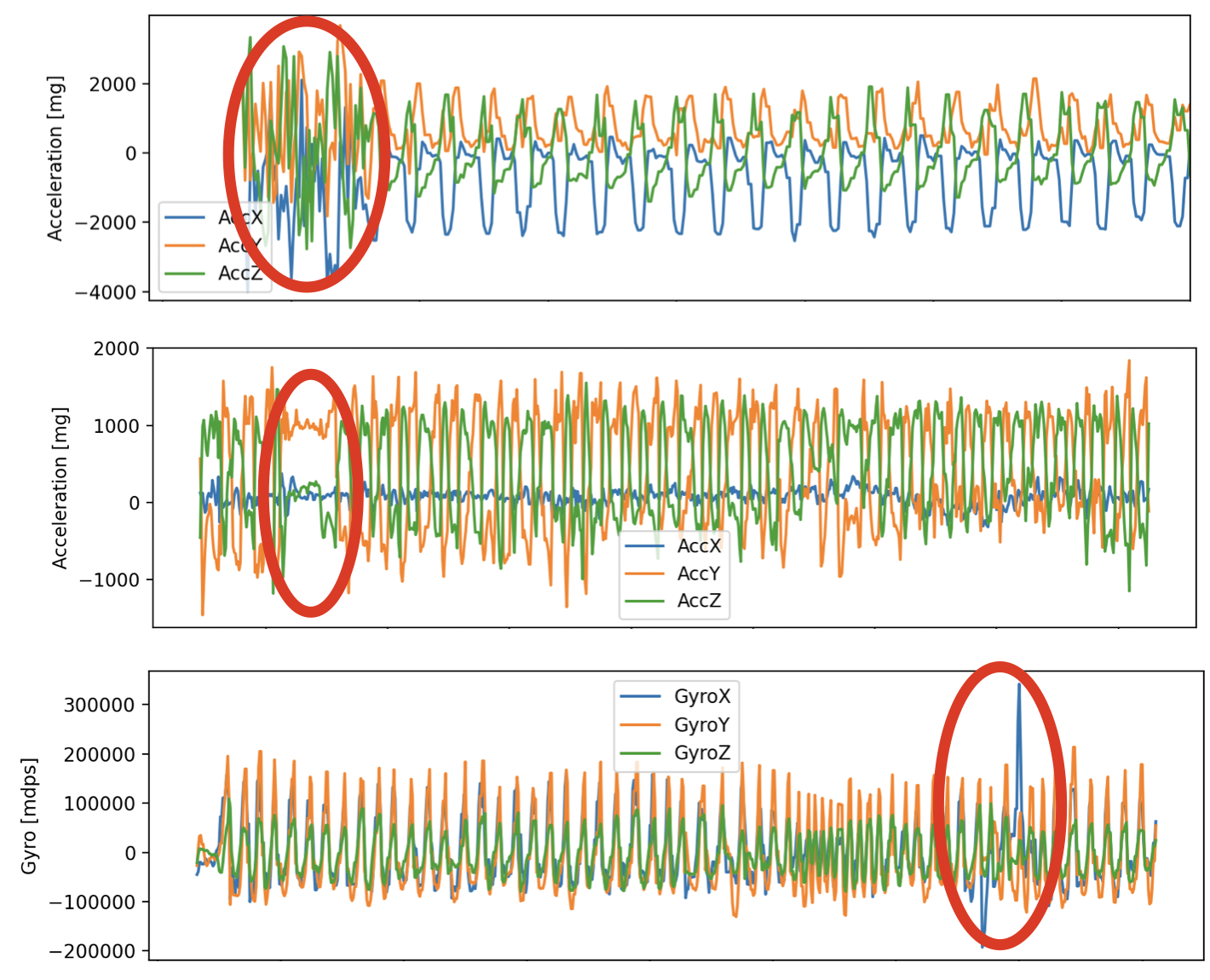}
    \caption{Example of Outlier/Noise instances}
    \label{fig:enter-label}
\end{figure}

\subsection{Data Downsampling}

In this study, all signals were downsampled from an original sampling rate of 104 Hz to 26 Hz. This downsampling approach served dual purposes: it reduced computational memory requirements by a factor of four and simultaneously enhanced the accuracy of the machine learning models by acting as a denoising mechanism. By decreasing the sampling rate, the process emphasized the global shape characteristics of the signals, which are critical for classification, over the finer, bit-by-bit details. This focus on overarching signal morphology aligns with findings this study on electrocardiogram (ECG) signals \cite{mepaper}, which have underscored the importance of overall shape rather than granular data points. Consequently, the downsampling methodology not only optimizes computational efficiency but also contributes to the robustness and accuracy of the models.

\begin{figure}[H]
    \centering
    \includegraphics[width=1\linewidth]{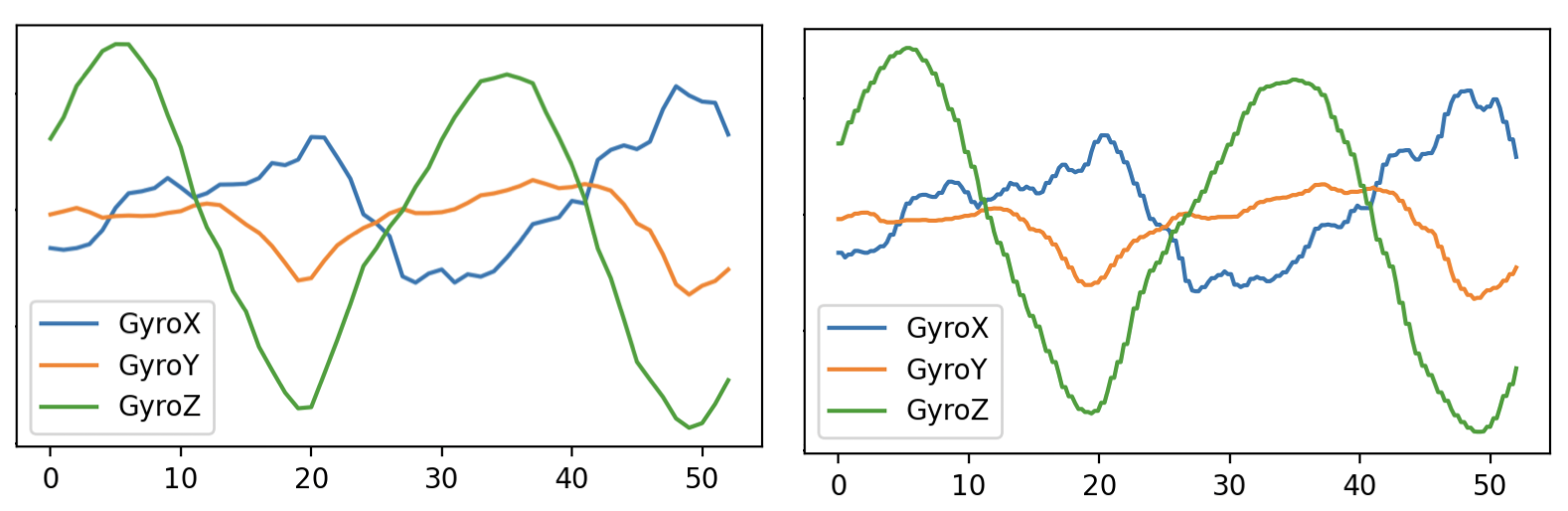}
    \caption{Example of Downsampling effect: Left hand side at 26Hz, Right hand side at 104Hz}
    \label{fig:enter-label}
\end{figure}

\subsection{Data Window}

\begin{figure}[H]
    \centering
    \includegraphics[width=1\linewidth]{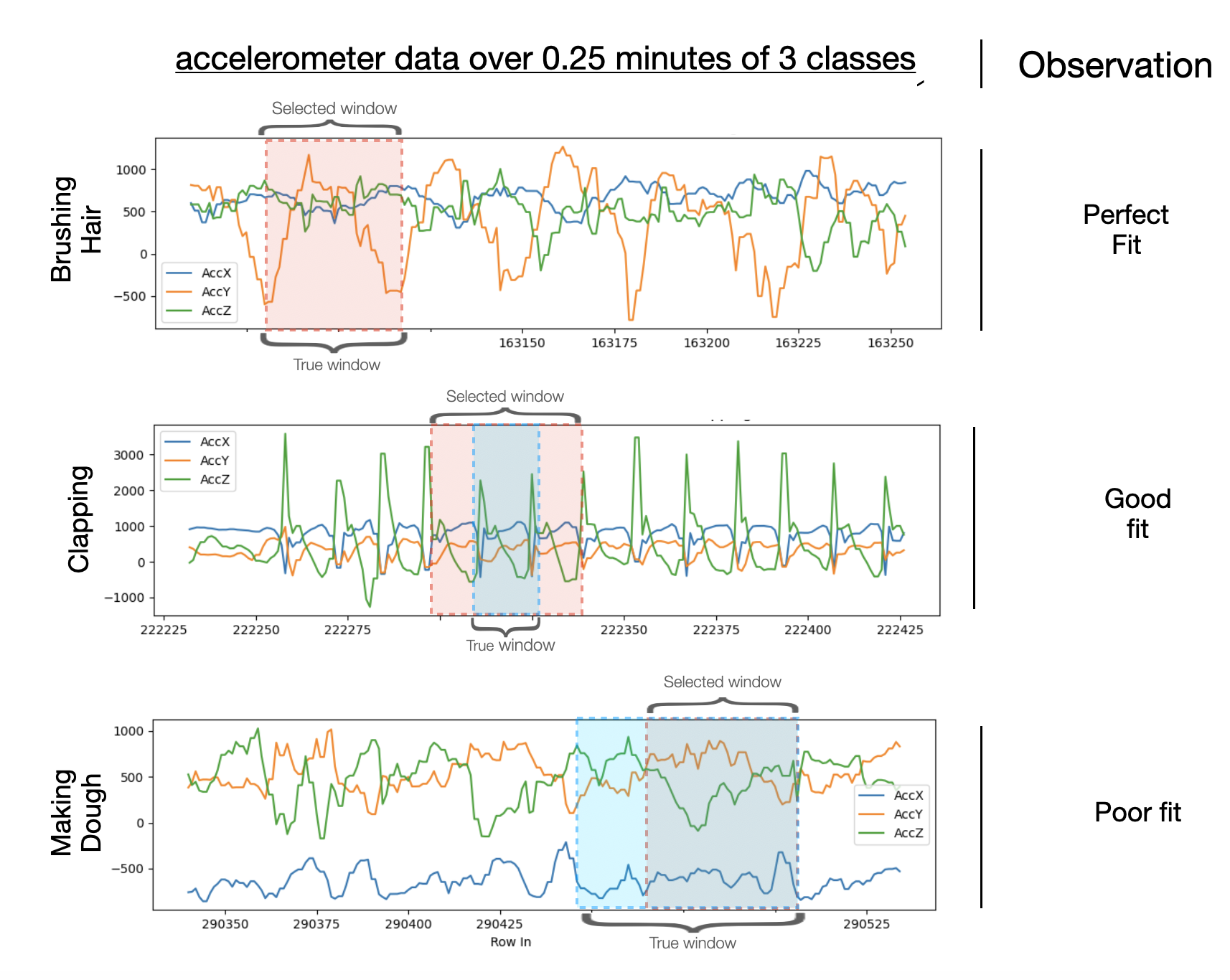}
    \caption{example of the effect of window size on different gestures}
    \label{fig:window}
\end{figure}

In determining the appropriate window size for data analysis, a duration of \textbf{1.5 seconds} at a sampling rate of 26 Hz was selected. This window size effectively represents the average consistent periodic movement observed across most classes in the dataset. The appearance of periodic movements within the data classes allows for accurate predictions within the 1.5-second window, as the model can effectively capture and analyze these recurring patterns. However, longer gestures, such as "Making Dough," posed challenges because the model was trained on segments of these movements rather than their full duration, limiting its ability to accurately predict extended actions as demonstrated in Fig \ref{fig:window}. Despite this limitation, the chosen window size ensures compatibility with short, periodic data patterns. The selected 1.5 seconds window size was calculated based on the average time of two consecutive major peaks from all data recordings.

\subsection{Unclassified Class}

The Unclassified class comprises signal behaviors that have the potential to cause misclassification during the training process. This class is constructed by incorporating trimmings and outliers derived from other predefined classes, as well as random high-velocity movements. The primary objectives of introducing the Unclassified class are to improve boundary decision-making and to enhance the overall robustness of the model during training. By including data that represents ambiguous or non-categorized motion, the model becomes better equipped to handle variability and uncertainty in real-world scenarios.

\subsection{Feature Extraction}

 Feature extraction \cite{features} is a typical process in signal processing, we selected a set of statistical features from each column of the raw sensor data, specifically AccX, AccY, AccZ, GyrX, GyrY, and GyrZ. The selection of these features was guided by the need to minimize computational complexity. Within each data window, 13 distinct features were computed for each of the six sensor axes, resulting in a comprehensive total of 78 features per window. This approach ensures that the feature set is both efficient and effective for subsequent machine learning tasks, facilitating the development of robust models within the TinyML framework by leveraging a balanced combination of computational feasibility and descriptive power.

\begin{itemize}
    \item \( \max(x) \): Maximum value
    \item \( \min(x) \): Minimum value
    \item \( \text{mean}(x) \): Mean value
    \item \( \text{std}(x) \): Standard deviation
    \item \( \text{range}(x) \): Range
    \item \( \text{absm}(x) \): Absolute mean
    \item Root Mean Square: \( \sqrt{\frac{1}{N} \sum_{i=1}^N x_i^2} \)
    \item P2P Low/High Frequency: \( \max(x_{\text{l/h}}) - \min(x_{\text{l/h}}) \)
\end{itemize}
\begin{itemize}
    \item Average Magnitude Difference: \( \frac{1}{N-1} \sum_{i=1}^{N-1} |x_{i+1} - x_i| \)
    \item Zero Crossing Rate: \( \frac{1}{N-1} \sum_{i=1}^{N-1} \mathbb{1}((x_i \cdot x_{i+1}) < 0) \)
    \item Mean Crossing Rate: \( \frac{1}{N-1} \sum_{i=1}^{N-1} \mathbb{1}(((x_i - \mu) \cdot (x_{i+1} - \mu)) < 0) \)
    \item Mean Absolute Deviation: \( \frac{1}{N} \sum_{i=1}^N |x_i - \mu| \)
\end{itemize}

\subsection{UMAP Classes Distribution}

UMAP (Uniform Manifold Approximation and Projection) \cite{UMAP} is a dimensionality reduction technique, similar to t-SNE \cite{tsne}, that visualizes high-dimensional data in low dimensions while preserving its structure. The UMAP distribution of data classes demonstrates defined clusters, which serves as a significant indicator for the success of the data preprocessing when compared to raw data, which exhibits random and unclustered distributions. This preprocessing not only preserves critical time series information but also eliminates the necessity for computationally intensive models, such as those based on Long Short-Term Memory (LSTM) \cite{lstm, lstmsignals} networks or attention mechanisms \cite{attention, transformer}. By maintaining clear and distinct clusters, the preprocessing steps ensure that machine learning models can operate with reduced computational complexity, leading to lower memory consumption and enhanced efficiency.

\begin{figure}[H]
    \centering
    \includegraphics[width=0.8\linewidth]{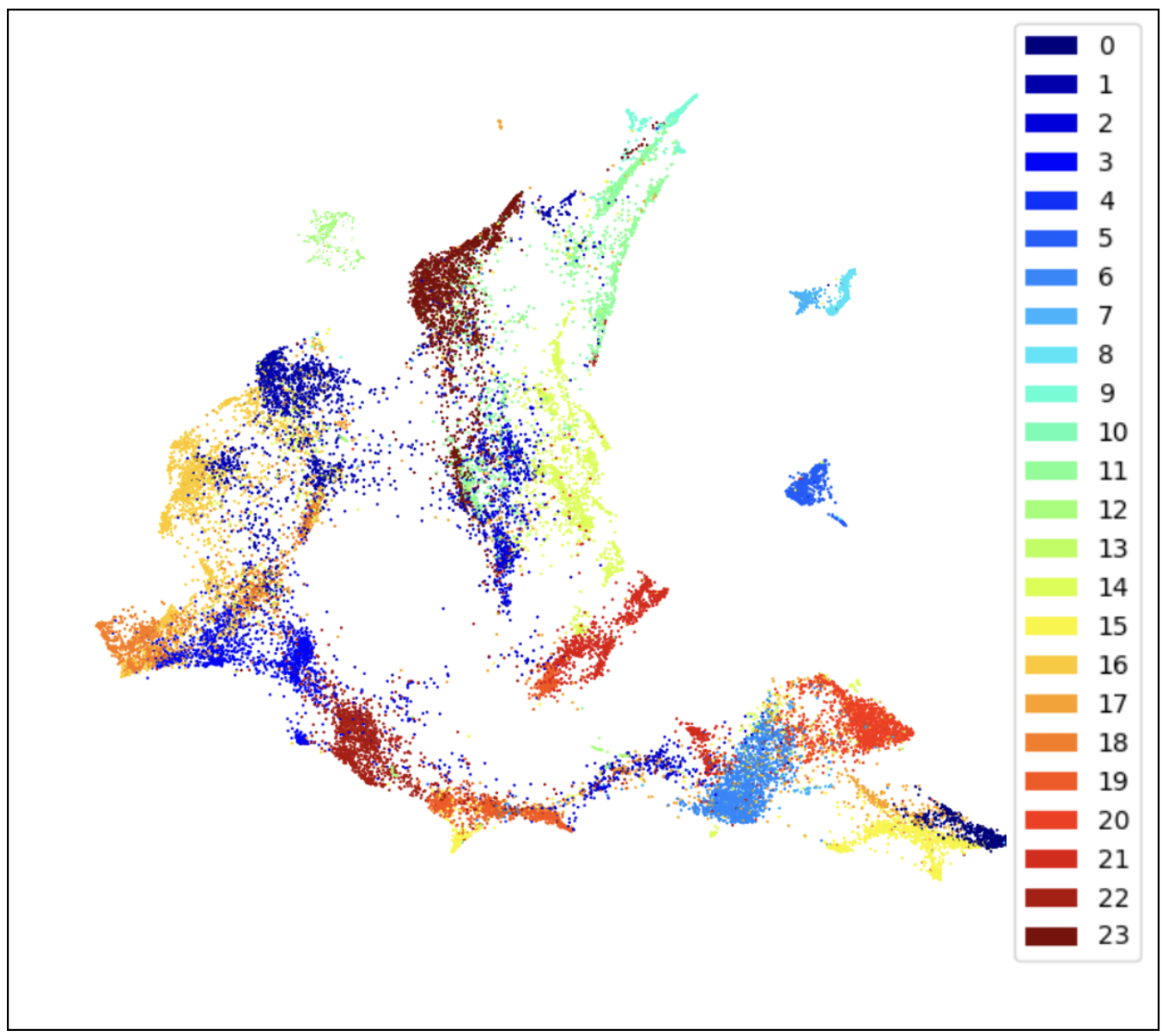}
    \caption{Umap distrubution of the 24 data classes}
    \label{fig:enter-label}
\end{figure}

\begin{table*}[tb]
  \centering
  \caption{Model Performance on Raw and Preprocessed Data}
  \label{tab:performance_comparison}
  \footnotesize{
    \begin{tabular}{l|lc|cccc}
      \toprule
      \multicolumn{2}{c}{\textbf{Model}} & \textbf{Open Source} & \textbf{Accuracy (\%)} & \textbf{Parameters} & \textbf{Stack Footprint}  & \textbf{Program Footprint} \\ 
      \midrule
      \multirow{3}{*}{\rotatebox[origin=c]{90}{\textbf{Raw}}}
        & LightGBM & Yes & 84.54 ± 1.02 & 5,389 & 2.5KB & 13Kb*** \\
        & XGBoost & Yes & 84.42 ± 1 & 5,389 & 2.3KB & 13Kb*** \\
        & \textbf{1D ResNet (Baseline)**} & Yes &  \textbf{97.01} & 1,071,832 & - & - \\
      \midrule
      \multirow{6}{*}{\rotatebox[origin=c]{90}{\textbf{Preprocessed}}}
        & 1D ResNet**         & Yes &  93.65 & 1,071,832 & - & - \\
        & LightGBM          & Yes &  86.23 ± 1.44 & 5,389 & 1.8KB & 13Kb*** \\
        & XGBoost           & Yes &  86.84 ± 1.24 & 5,389 & 1.7KB & 13Kb*** \\
        & \textbf{Basic NN**}          & Yes &  \textbf{92.47 ± 0.55} & 7,928 & 4.2Kb & 21Kb*** \\
        & Neuton AutoML     & \textbf{No} &  91.15 & 1,813 & 2KB & 17Kb \\
        & \textbf{XGBoost\textsuperscript{*}} & Yes & \textbf{84.7} & 5,389 & \textbf{0.82KB} & 13Kb*** \\
      \bottomrule
    \end{tabular}
    
    \vspace{0.3em} 

    \begin{minipage}[t]{0.3\textwidth}
      \footnotesize{\textsuperscript{*} The model was trained using the top 20\% most important features, as determined from the feature importance matrix of a model trained on the entire dataset. The features are:
      \texttt{
        ACC\_Y\_STD\_DEV, ACC\_X\_MAX, ACC\_X\_MEAN, ACC\_X\_MIN, ACC\_Z\_MEAN, ACC\_Z\_MIN, ACC\_Y\_MAD, ACC\_X\_RMS, ACC\_Z\_P2P\_HF, ACC\_Y\_P2P\_LF, ACC\_Z\_RMS, ACC\_Y\_MIN, GYRO\_Y\_MAD, ACC\_Y\_ZCR, ACC\_Y\_P2P\_HF, GYRO\_Y\_AMDF
      }}
    \end{minipage}
    \hfill
    \begin{minipage}[t]{0.3\textwidth}
      \footnotesize{\textsuperscript{**} The model was trained using the early stopping technique with a patience of 30 epochs and a batch size of 1,024 on an 8GB GPU Nvidia 1080. An initial learning rate of 3 × $10^{-4}$, known as Karpathy's constant \cite{andrej}, was employed with learning rate decay. Training optimization was performed using the AdaBelief \cite{adabelief} optimizer.}
    \end{minipage}
    \hfill
    \begin{minipage}[t]{0.3\textwidth}
      \footnotesize{\textsuperscript{***} 
For decision tree-based models, converting the model into if-else statements uses excessive program memory because weights are stored as strings in the generated C code. To optimize memory usage, we convert the model into a serialized binary format and use a custom tree traversal function. For neural network-based models, we store the weights similarly, with a custom feedforward function, both written in C.
      }
    \end{minipage}
    \\
  }
\end{table*}

\section{Expirements}
\label{inject}
To scale the solution to 24 classes while maintaining memory usage below 2 KB, we implemented an incremental class injection strategy. This process involves sequentially adding classes to the model, enabling it to adapt to new data through incremental training. After each training phase, the updated model is deployed in an inference setup, where \textbf{manual evaluation is conducted} to assess potential overlap between the newly added class and the existing classes. If the overlap extends to more than one existing class, the new class is discarded to prevent confusion and maintain clear class boundaries. For overlaps with a single class, semantic analysis is performed to determine whether the classes can be merged. If merging is feasible, the classes are combined to maintain a cohesive structure; otherwise, the newly added class is excluded. Without this strategy, the system's capacity was limited to handling 12–16 classes. Using this approach, we scaled the model to 24 distinct classes, selected from an initial set of 31 gestures. This selection process was supported by the Neuton AutoML platform, where multiple experiments were conducted to refine and validate the final set of classes. This combination of incremental class injection and overlap management ensures the model's scalability while adhering to strict memory constraints, making it effective for real-world deployment.

After obtaining the final set of classes, we train and evaluate models on both raw and preprocessed data using the following architectures:

\begin{itemize}
    \item Expirement 1: 1D ResNet, We adopt the ResNet architecture \cite{resnet} as implemented in \cite{mepaper}.
    \item Expirement 2: Gradient Boosting Models: We utilize two gradient boosting architectures, XGBoost \cite{xgb} and LightGBM \cite{lightgbm}.
    \item Expirement 3: Basic Neural Network: A simple feedforward neural network with three fully connected layers consisting of 64, 32, and 24 neurons, respectively. Each layer is followed by ReLU activation, with dropout (rate 0.2) applied after the first two layers to mitigate overfitting. ReLU was chosen for ease of implementation on the ISPU instruction set and for its improved latency compared to other activations, such as softmax.
\end{itemize}

\section{Results}

The purpose of the first experiment is to establish a benchmark for evaluating TinyML models. For this purpose, we use a 1D ResNet as the baseline model, which achieves the highest accuracy of 97.01\% on raw data. This superior performance is expected, as the convolutional \cite{cnn1, cnn2} layers in ResNet effectively learn to extract complex features directly from the raw input, outperforming models that rely solely on pre-extracted statistical features. However, this accuracy comes at a significant cost, with the model containing over 1 million parameters. When trained on preprocessed data, the model's accuracy stagnates at 93.65\%. This suggests that the performance plateau for this dataset is approximately aligned with the performance of models trained on the pre-extracted features. The inability of the model to surpass this threshold indicates that the pre-extracted features do not contain enough additional information for the model to learn more advanced representations. This highlights a limitation in feature extraction for this specific dataset.

For gradient boosting models such as LightGBM and XGBoost in expirement 2, we had to reduce the model size to comply with the 32KB memory limit of the ISPU, which negatively impacted their performance. These models lack native feature extraction capabilities, making them more effective when provided with well-engineered features. The training results demonstrated a 2\% improvement in accuracy when using preprocessed data compared to raw data, as observed across a 5-fold cross-validation. This underscores the effectiveness of the feature extraction process, which is further validated by the clearly defined clusters in the UMAP distribution plot. 

In experement 3, With a memory footprint comparable to other approaches, we trained a basic neural network architecture that achieved state-of-the-art performance of 92.47\% on the preprocessed data. This result is highly competitive and aligns with the constraints of the ISPU platform, making it a viable solution. Notably, this performance surpasses that of the closed-source Neuton AutoML platform.

Additionally, to further reduce stack memory consumption, we retrained the XGBoost model using only the top 20\% most important features, as determined by the feature importance matrix from the previous training. This approach maintained relatively the same performance while significantly reducing stack memory usage to just 0.8KB. This optimization highlights the potential for balancing model performance and resource constraints on platforms with strict memory limitations.
The results of this study demonstrate the efficacy and efficiency of the proposed model in accurately classifying 24 distinct classes. The set of optimally generated models operating with minimal resource requirements, utilizing only 2KB of data memory or less and 22 KB of program memory at maximum, which underscores the suitability for deployment in The ISPU. 

\section{Discussion}

While overall accuracy is a crucial metric for evaluating model performance, it does not fully capture all aspects of a model's operational effectiveness. In our study, we observed that the model consistently and correctly classified instances of a any class as long as there is a high correlation between the rhythmic periods of movements within that class, as illustrated in Figure 4.
\begin{figure}[H]
    \centering
    \includegraphics[width=1\linewidth]{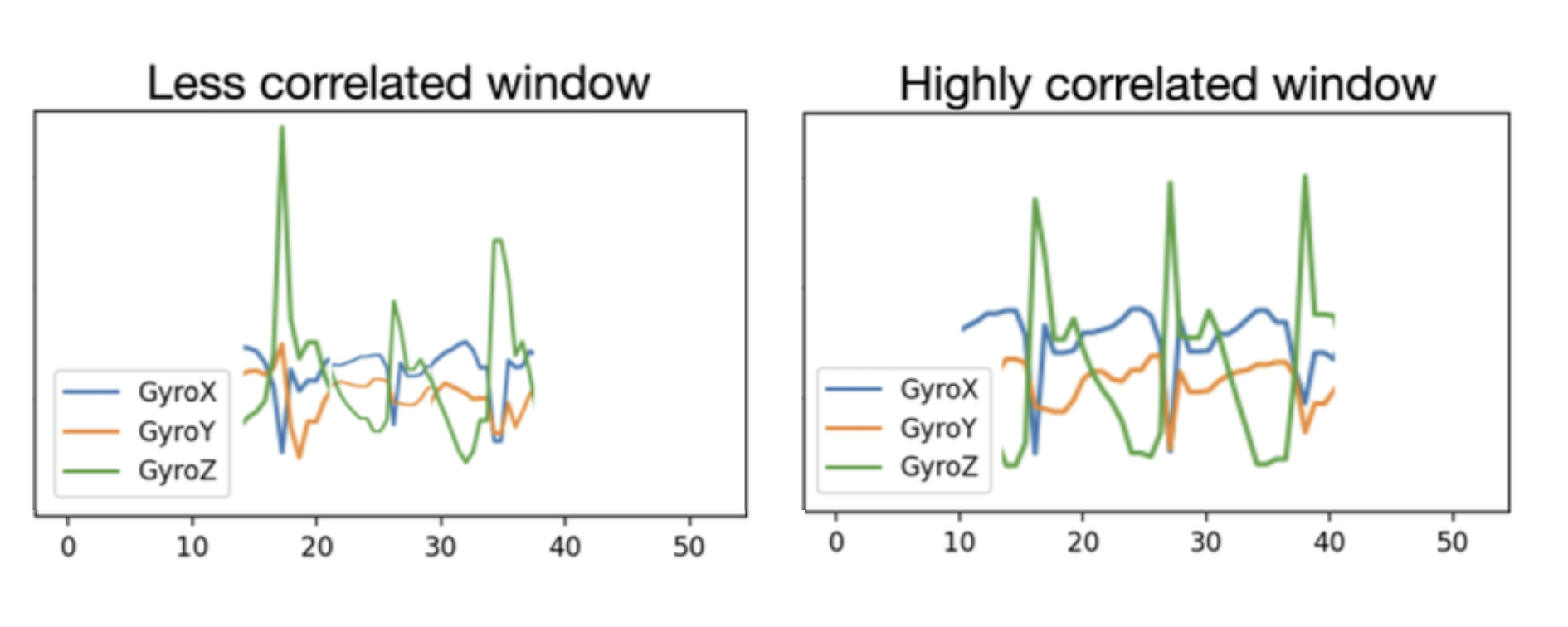}
    \caption{Visual Representation of two examples of data windows self correlation}
    \label{fig:enter-label}
\end{figure}
This occurred regardless of the model's overall accuracy, highlighting the model's ability to reliably identify specific movement patterns with strong temporal consistency.
This emphasizes the importance of evaluating not only the model's accuracy but also its consistency and robustness in classification under varying real-world conditions.

During the incremental class injection process, certain classes were either discarded or merged to enhance model performance and maintain memory efficiency. The "Brushing Teeth" class was discarded due to a high incidence of false positives with other classes within the cleaning category, such as "Brushing Hair," "Washing Face," and "Shaving." This misclassification was primarily attributed to the need for specific gyroscope calibration, as the high-frequency motions involved in brushing teeth differed significantly from those of other cleaning activities. Additionally, the "Driving" class was excluded because it lacked periodic movement within the selected window size, and the random acceleration and braking associated with driving introduced unpredictable spikes in the data. Furthermore, when a vehicle was moving at a relatively constant speed on a straight road, the system erroneously classified it as “handstill” based on the principle of inertia. On the other hand, certain classes were merged to improve classification stability and reduce confusion. For instance, the "Writing" class exhibited misclassifications between return movements and forward movements, which were resolved by combining them into a single class similarly with "Writing on Board" class. The "Using Computer" class also benefited from merging, as it initially caused the model to switch randomly between two distinct classes “using keyboard” and “ using mouse”. By consolidating these classes, the model's reliability and accuracy were significantly improved. These adjustments—discarding problematic classes and merging similar ones—were crucial in maintaining a cohesive and efficient class structure. 

The tradeoff between accuracy and the number of classes highlights the model's scalability, as it maintains high accuracy even with an increasing number of classification categories. While expanding the number of classes typically reduces accuracy due to increased complexity and class overlap, this study demonstrates robust performance through an incremental class injection strategy that minimizes class confusion. This balance validates the model’s adaptability and efficiency especially for neural network based models, making it well-suited for resource-constrained TinyML applications and dynamic environments requiring complex classification.

\section{Conclusion}
This study demonstrates the feasibility of deploying HAR models directly on the ISPU, overcoming the stringent memory and computational constraints of these ultra-constrained devices. We developed a 24-class HAR model achieving 85\% accuracy, using less than 850 bytes of stack memory. Our tailored data preprocessing pipeline and training strategy ensure high efficiency while maintaining a minimal memory footprint.
The results highlight the transformative potential of on-sensor processing for applications such as smart homes, healthcare monitoring, and automation ,bearing fault detection and predictive maintenance, particularly in scenarios requiring low latency and enhanced privacy. By enabling on-sensor processing, our approach enhances real-time capabilities, data privacy, and energy efficiency. The release of a diverse HAR dataset with 24 gestures further supports advancements in this field. While our study focuses on HAR, the techniques and methodologies introduced are broadly applicable to other edge computing tasks on ISPUs.
Future directions include exploring advanced model compression and pruning techniques,  and integrating ISPUs into larger IoT ecosystems, paving the way for scalable, decentralized intelligence at the network's edge. This work underscores the critical role of combining hardware innovations with machine learning to realize the potential of edge computing and TinyML.

\newpage
\newpage
\bibliographystyle{ACM-Reference-Format}
\bibliography{library}

\appendix

\end{document}